\documentstyle[11pt,paspconf]{article}
\begin{document}
\title{The Angular Two Point Correlation Function for the FIRST Radio Survey}

\author{C.M.Cress, D.J.Helfand}
\affil{Department of Astronomy, Columbia University, 538 W120th St, New York,
 NY 10027}
\author{R.H.Becker\altaffilmark{1}, M.D.Gregg}
\affil{LLNL, Livermore, CA 94450}
\author{R.L.White}
\affil{STScI, 3700 San Martin Drive, Baltimore, MD 21218}
\altaffiltext{1}{Department of Physics, University of California, Davis, CA 95
616}
\thispagestyle{empty}
\begin{abstract}
The angular two-point correlation function is calculated for the first 300
square degrees of the FIRST radio survey. Results for various subsamples are
also obtained. Double-lobed sources are shown to have a higher clustering
amplitude than the sample as a whole. Small differences in the correlation
function from one region of the sample to another and results of various
flux cuts are discussed.

\end{abstract}
\pagestyle{empty}

\section{Introduction}

The first 300 square degrees of the FIRST (Faint Images of the Radio Sky at
Twenty centimeters) survey covers a strip from 07$^{h}$15 to 16$^{h}$30 in RA
and from about 28.4$^{\circ}$ to 31$^{\circ}$ in dec. This yields a catalog
of about 27,000 sources with positional accuracies better than
$1^{\prime\prime}$. About 30\% of these sources are in double or
multi-component
systems. Becker, White and Helfand (1995) estimate that the catalog is 80\%
complete to 1.0 mJy and 95\% complete to 2.0 mJy.

While some evidence for the clustering of bright radio sources ($>0.5$ Jy) has
been presented (e.g. Peacock \& Nicholson, 1991),
little is known about the clustering of fainter populations. FIRST provides
an excellent opportunity to investigate this problem.

\section{Determining the correlation function}
Landy and Szalay (1993) (LS) discuss four methods for
estimating the correlation function, $\omega(\theta)$.
To make an independent estimate of the
uncertainties associated with each method, we used the following procedure:
 16 random fields (R) with surface densities equal
to that of the survey (D) were generated using the random number generator `ra
n1'
given in Press et al. (1986). The correlation function was determined for each
 of the random fields using all four of the methods. For each method and for
each
$\theta$, then, there were 16 estimates of $\omega(\theta)$ distributed around
 zero.
 The standard deviation of
this distribution was taken as an estimate of the uncertainty in the
$\omega(\theta)$ calculated for the survey. In qualitative agreement
with LS, we find that on large
angular scales ($>10^{\circ}$) their methods (iii) and (iv) have smaller
associated
uncertainties.  On smaller scales, we find the uncertainties for the four
methods to be similar.

The correlation function for the whole sample is calculated using both
methods (ii) ($\omega(\theta)=DD/DR-1$) and (iii) ($\omega(\theta)=
(DD-2DR+RR)/RR$).

To avoid over-counting double-lobed radio galaxies and other
multi component
systems, sources separated by less than $0.02^{\circ}$ were counted as a
single source. This yields about 22,000 sources for correlation
function analysis. The random fields were not corrected for the bias resulting
from this procedure; that is, random sources separated by less than
 $0.02^{\circ}$ were
included in the random catalogs.

\section{Results}
\subsection{Whole Strip}
The results for all the sources in the sample are shown in Figure I.
Error bars are shown for method (iii) calculations only. They are obtained
from the random field process described above. Out
to $1^{\circ}$ or $2^{\circ}$ the correlation function clearly shows power
law behaviour. Fitting a function of the form
$\omega(\theta)=A\theta^{\gamma}$ out to $1.5^{\circ}$
yields A=0.012$\pm$0.001, $\gamma$=-0.88$\pm$0.05 for method (ii)
and A=0.007$\pm$0.001, $\gamma$=-1.03$\pm$0.1 for method (iii).
These slopes are somewhat larger than those determined for optical surveys
(a value of -0.8 is often taken as a standard estimate which is consistent
with our results at the 2$\sigma$ level)
It's possible that our uncertainties are underestimated enough for the
results to be consistent. The first FIRST catalogue does contain a few bad
fields
and some widely spaced doubles could still be in the sample. These would both
push up the counts in the small angle bins, increasing the slope somewhat.
However, Figure I does not really show any evidence for this.

There is an obvious flattening of the curve just past
1$^{\circ}$. It should be pointed out that this is half the width
of the survey strip:
analysis of the larger portion of the survey will hopefully reveal whether
this flattening is physically significant. The other obvious feature
is the
drop off near $20^{\circ}$. The correlation
between the position of this cutoff and the size of the sample can be
tested when the
extended catalogue becomes available.

We intend to use estimates of the radio
luminosity function to infer information about the spatial correlation
function. This will allow a more meaningful comparison of the correlation
amplitude with those obtained for other samples.

\subsection{Different regions of sky}
The strip was divided into four portions and the correlation function was
determined using method (ii) for each quarter. There is some evidence for
variation in the slope and amplitude of the
correlation function amongst these different $2.5 \times 34^{\circ}$ cells,
although the larger sample will be needed before a quantitative statement can
be made.

There appears to be some correlation between the position of Abell clusters
and
the strength of the correlation signal. In an attempt to establish
how much the radio sources in Abell clusters contribute to the correlation
function,
sources within 0.4$^{\circ}$ of the center of all Abell clusters were excluded
from the calculation. We found that the surface density of sources was indeed
larger
than average within $0.4^{\circ}$ of clusters, but that removing these sources
 did not
significantly affect the correlation function (except to increase the
uncertainties).
Section 3 ($11^{h}48$ to $14^{h}00$ in RA) appears to have the highest
correlation amplitude. This could be
related to the presence of nearby clusters such as Coma in this region.

\subsection{Correlation of Double and Multi-Component Sources}
A catalogue of 3800 double-lobed and multi-component sources was created by
collapsing all
sources within 0.02$^{\circ}$ of each other to a single source. The angular
correlation
function for these sources (determined using method (ii)) is shown in
Figure II. The
best fit to the doubles correlation function yields A=0.032$\pm$0.003,
$\gamma$=
-1.10$\pm$0.05.
The large correlation seen could be explained if these resolved doubles were
 on average
closer to us than the remainder of the sample.

We also analysed a sample which included all sources $except$ double and
multi-component systems.
The best fit yielded A=-0.01$\pm$0.001, $\gamma$=-0.76$\pm$
0.05, more consistent, in slope, with optical survey results.

\subsection{Flux Cuts}
The correlation function was also determined for samples where all
sources below a certain flux limit were omitted. The samples with 2 mJy,
 4 mJy and 10 mJy cuts showed similar correlation functions (although the
scatter increased as the number of sources decreased). At 1.0 mJy
the expected
ratio of the number of starburst galaxies to the
number of AGN's is of order unity. Above 2 mJy this ratio becomes orders of
magnitude larger (Condon 1991). It appears that removing these relatively
nearby starburst galaxies does not result in a significant loss of signal.
This provides some evidence that we are finding structure in more distant
populations.

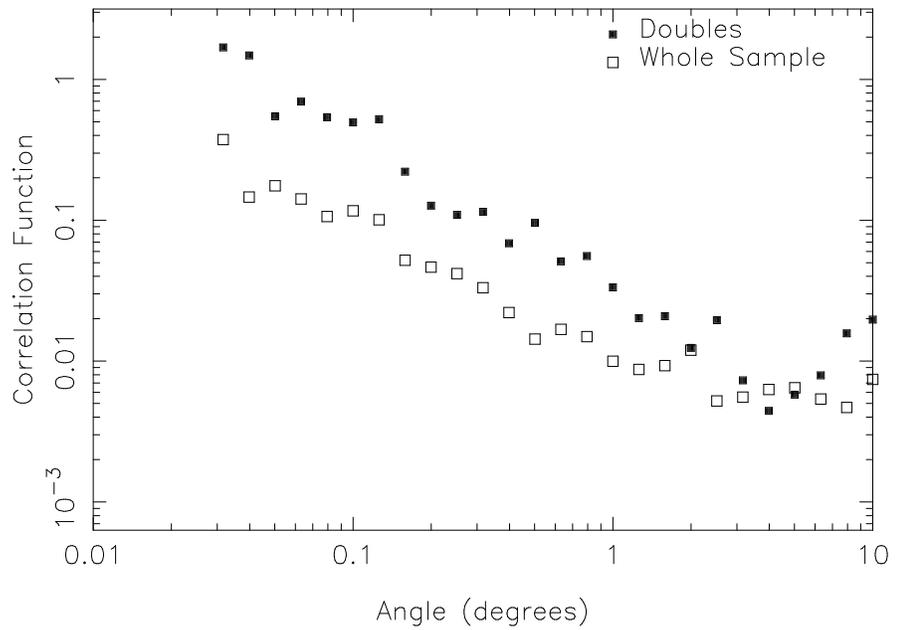
\end{document}